\pdfoutput=1
\documentclass{article}
\usepackage{spconf,amsmath,graphicx}

\usepackage{enumitem}
\setlist{nosep, leftmargin=14pt}

\usepackage{censor}

\usepackage{mwe} % to get dummy images
\newlength{\tempdima}
\newcommand{\rowname}[1]% #1 = text
{\rotatebox{90}{\makebox[\tempdima][c]{#1}}}
\usepackage{graphbox}
\usepackage[colorlinks=true,linkcolor=red, citecolor=blue]{hyperref}%
\usepackage[super]{nth}
\usepackage{amssymb}
\usepackage{mathtools}
\usepackage{multirow}
\usepackage{caption}
\usepackage{subcaption}
\usepackage{dblfloatfix}
\usepackage{array}
\usepackage{booktabs}
\usepackage{siunitx}
\usepackage[export]{adjustbox}
\newlength{\wdth}

\usepackage[mincitenames=1,minbibnames=1,maxcitenames=2,maxbibnames=2,style=ieee]{biblatex}

\newcolumntype{P}[1]{>{\centering\arraybackslash}p{#1}}
\title{Towards Measuring Domain Shift in Histopathological Stain Translation in an Unsupervised Manner}

\name{Zeeshan Nisar$^{\star}$ \quad Jelica Vasiljević$^{\star,\dagger,\text{\textbardbl}}$ \quad Pierre Gan\c{c}arski$^{\star}$  \quad Thomas Lampert$^{\star}$}
\address{$^{\star}$ICube, University of Strasbourg, France \quad 
		 $^{\dagger}$University of Belgrade, Serbia \\
		 $^{\text{\textbardbl}}$Faculty of Science, University of Kragujevac, Serbia}

\begin{document}
%\ninept
%
\maketitle
\begin{abstract}
%One of the greatest challenges with deep neural networks is generalisation. A network trained on source data may not generalise well, particularly when domain shift is introduced. 
%In digital histpathology, this domain shift can occur when different stains are used, as well as when translating stains, using different scanners, etc.
%%This can occur in digital histopathology with different stains, when translating stains, different scanners, inter-centre variation, etc.
%%and performs significantly worse when applied to unseen target data but for same task. This significant drop in performance is observed because of domain shift. 
%An important step towards being robust to domain shift is the ability to detect and measure it. This article investigates the PixelCNN and domain shift metric to detect and quantify domain shift in digital histopathology.
%%both visually and quantitatively. 
%%In our initial experiments we performed a study on medical imaging (specifically on digital pathology) but this is a general solution and is applicable to all computer vision applications observing a domain shift.
%Moreover, this paves the way for a mechanism to infer the average performance of a model (trained on source data) on unseen and unlabelled target data. %is going to perform on an unseen target data without having any expert opinion or ground-truths.

Domain shift in digital histopathology can occur when different stains or scanners are used, during stain translation, etc. A deep neural network trained on source data may not generalise well to data that has undergone some domain shift.
An important step towards being robust to domain shift is the ability to detect and measure it. This article demonstrates that the PixelCNN and domain shift metric can be used to detect and quantify domain shift in digital histopathology, and they demonstrate a strong correlation with generalisation performance.
These findings pave the way for a mechanism to infer the average performance of a model (trained on source data) on unseen and unlabelled target data.
\end{abstract}
\section{Introduction}
\label{sec:intro}
Image datasets in digital pathology often consist of consecutive tissue slides stained differently \cite{zee1vasiljevic2021towards}, with each stain providing different information on the same region of interest. Since each stain highlights different tissue structures, even consecutive slides (representing identical anatomical structures e.g.\ glomeruli) can appear very different, see Fig.\ \ref{fig:cyclegan_translations} (\nth{1} row). Furthermore, the staining procedure is vulnerable to high variability due to inter-subject variations, lab specific techniques, capturing pipeline changes \cite{zee20stacke2020measuring}, scanner characteristics, and staining protocols, and these can introduce further variation of a tissue's appearance \cite{zee2leo2016evaluating}. 
%Data collected from the same medical centers and with the same scanner can also experience a domain shift because of changes in the capturing pipeline over time \cite{zee20stacke2020measuring}. 

Although large-scale biological structures retain morphological structure across each stain as with the glomeruli in Fig.\ \ref{fig:cyclegan_translations}, state-of-art deep learning (DL) methods trained for some task (i.e.\ glomeruli segmentation) on one (source) stain (e.g.\ PAS) do not generalise well to histological images of the target stain (e.g.\ Jones H\&E, CD68, CD34, Sirius Red), see Table \ref{table:MDS1} (\nth{1} row). This degradation in performance is caused by the inter-stain variation, see Fig.\ \ref{fig:cyclegan_translations} (\nth{1} row), or intra-stain variation (e.g.\ one stain collected from different laboratories). Even small domain shifts may cause significant drops in performance. Many DL algorithms are vulnerable to this shift \cite{zee18csurka2017comprehensive}, meaning that proper care needs to be taken when deploying them for clinical aid. As such, it is of great importance to handle this variance or at least to estimate when it is likely to significantly effect an algorithm's performance.

Therefore an important step towards handling domain shift in digital histopathology is the ability to detect it. To the best of our knowledge, no such work exists for digital pathology, particularly for segmentation. This article concentrates on detecting domain shift during stain style transfer between a source stain (PAS) and translated Target$\rightarrow$PAS, nevertheless the presented solution is general and unsupervised, and is therefore applicable to other types of domain shift. 

Two approaches are investigated: the PixelCNN \cite{zee19salimans2017pixelcnn++}, which is used to model the distribution of the source data in an unsupervised manner, and the Domain Shift Metric (DSM) \cite{zee20stacke2020measuring}, which uses a pre-trained feature representation to measure domain shift and therefore integrates some knowledge of the task to be performed, in this study the segmentation representation trained on the source stain is used. It is shown that both of these measures have high correlation with whole slide image (WSI) segmentation scores, even though the domain shift is calculated on a small subset of the data. Except for pre-training the segmentation model in DSM, both of these approaches measure domain shift in an unsupervised manner.
The rest of this paper is organised as follows: Section \ref{sec:litrev} reviews the literature on stain translation and how it can introduce domain shift, Section \ref{sec:methods} details the methods used. Section \ref{sec:experiments_and_results} presents the dataset, experiments, and results. Finally, Section \ref{sec:conclusion_future} concludes the paper with possible future directions.

% % \vspace*{-3.0mm}
% \begin{figure}[!h]
% \begin{center}
% \setlength{\tabcolsep}{3.15pt} % Default value: 6pt
% \begin{tabular}{ccccc}
%     \small{PAS} & \small{Jones H\&E} & \small{CD68} & \small{Sirius Red} & \small{CD34} \\
% 	\includegraphics[width=1.5cm]{images/stain_images/02.png} & 
% 	\includegraphics[width=1.5cm]{images/stain_images/03.png} &
% 	\includegraphics[width=1.5cm]{images/stain_images/16.png} &
% 	\includegraphics[width=1.5cm]{images/stain_images/32.png} & 
% 	\includegraphics[width=1.5cm]{images/stain_images/39.png} 
% 	\vspace*{-3.5mm}
% \end{tabular}
% \caption{Different Staining methods containing glomeruli as RoI.}
% \label{fig:stains}
% \end{center}
% \end{figure}

\section{Literature Review}
\label{sec:litrev}

%zee4bentaieb2017adversarial
%zee5lafarge2017domain,
%zee7macenko2009method,
Though several approaches \cite{zee3de2019stain, zee6lafarge2019learning, zee8otalora2019staining, zee9shaban2019staingan} deal with the problem of intra-stain variation, few address the problem of inter-stain variation \cite{zee1vasiljevic2021towards, zee10lampert2019strategies, zee11gadermayr2019generative}. Stain colour augmentation, stain normalisation, and stain transfer \cite{zee12srinidhi2020deep} are the current standard approaches to learn stain invariant representations. The term stain invariant indicates the ability of the same model to be applied across multiple stains (possibly to those that are not used during training). In general, it is assumed that annotations for the same task are available for the source stain but not for the target stains since acquiring labels for each staining is expensive and laborious.  \textcite{zee13tellez2019quantifying} state that stain colour augmentation has a greater influence on the robustness of DL methods than stain normalisation. While stain transfer (a technique for virtual staining) tackles the generic problem of lack of annotations in the medical domain \cite{zee14lahiani2019virtualization, zee15mercan2020virtual} and can be applicable to various related scenarios. Recently \textcite{zee11gadermayr2019generative} propose to use an unpaired adversarial image-to-image translation approach called CycleGAN \cite{zee16zhu2017unpaired} to overcome the lack of annotations for the target stain by: 1) training a segmentation model on the source stain and apply it to the target stain translated to the source stain, named as MultiDomain Supervised 1 (MDS1); 2) translate the source to target stain and directly train the segmentation model for the target stain, named as MultiDomain Supervised 2 (MDS2). \textcite{zee1vasiljevic2021towards} extend this to create UDAGAN, a stain augmentation procedure that uses multiple target stains to learn a stain invariant representation for segmentation, which can even be applied to out-of-sample stains.

\begin{table}[]
\centering
%\small
\captionsetup{}
     \begin{tabular}{P{1.4cm} P{0.8cm} P{1cm} P{0.8cm} P{0.8cm} P{0.8cm}}
      \hline 
      \multirow{3}{*}{\shortstack{Training\\ Strategy}} & \multicolumn{5}{c}{Test Stain}\\
      & \multirow{2}{*}{PAS} & \multirow{2}{*}{\shortstack{Jones \\H\&E}} & \multirow{2}{*}{CD68} & \multirow{2}{*}{\shortstack{Sirius \\Red}} & \multirow{2}{*}{CD34}\\
      & & & & &  \\
      \hline
      \hline
      \multirow{3}{*}{\shortstack{Baseline \\ PAS \\ \footnotesize{(Full slide)}}}  & \multirow{3}{*}{\shortstack{0.907 \\ \scriptsize{(0.009)}}} & \multirow{3}{*}{\shortstack{0.084 \\ \scriptsize{(0.033)}}} & \multirow{3}{*}{\shortstack{0.001 \\ \scriptsize{(0.001)}}} & \multirow{3}{*}{\shortstack{0.016 \\ \scriptsize{(0.018)}}} & \multirow{3}{*}{\shortstack{0.070 \\ \scriptsize{(0.063)}}}\\
      & & & &  \\
      & & & &  \\      
      \hline
      \multirow{3}{*}{\shortstack{MDS1 \\ \footnotesize{Target$\rightarrow$PAS} \\ \footnotesize{(Full slide)}}}  & \multirow{3}{*}{\shortstack{- \\ \scriptsize{(-)}}} & \multirow{3}{*}{\shortstack{0.849 \\ \scriptsize{(0.017)}}} & \multirow{3}{*}{\shortstack{0.683 \\ \scriptsize{(0.043)}}} & \multirow{3}{*}{\shortstack{0.870 \\ \scriptsize{(0.009)}}} & \multirow{3}{*}{\shortstack{0.754 \\ \scriptsize{(0.008)}}}\\
      & & & &  \\
      & & & &  \\
      \hline
    %   \multirow{3}{*}{\shortstack{MDS1 \\ \footnotesize{Target$\rightarrow$PAS} \\ \footnotesize{(2000 patches)}}}  & \multirow{3}{*}{\shortstack{- \\ \small{(-)}}} & \multirow{3}{*}{\shortstack{0.934 \\ \small{(0.0008)}}} & \multirow{3}{*}{\shortstack{0.713 \\ \small{(0.061)}}} & \multirow{3}{*}{\shortstack{0.944 \\ \small{(0.003)}}} & \multirow{3}{*}{\shortstack{0.788 \\ \small{(0.009)}}}\\
    %   & & & & & \\
    %   & & & & &  \\
    %   \hline
     \end{tabular}
     \caption{
     Segmentation scores (F$_1$) of the U-Net (trained on PAS) %of the full test slides and 2000 test patches (1000 glomeruli and 1000 negative)  
     applied to full test slides of different stains (\nth{1} row) and 
     translated (Target$\rightarrow$PAS) slides (\nth{2} row). Averages of 5 U-Net repetitions applied to 3 Cycle-GAN repetitions, i.e.\ 15 repetitions in total, standard deviations are in parentheses.
     %Larger the F$_1$ score smaller will be the $R_l$ distance between two distributions.
     }
    \label{table:MDS1}
\end{table}

\begin{figure}[]
\setlength{\tabcolsep}{1.5pt}
\captionsetup{}
    \small{
    \settoheight{\tempdima}{\includegraphics[width=1.5cm]{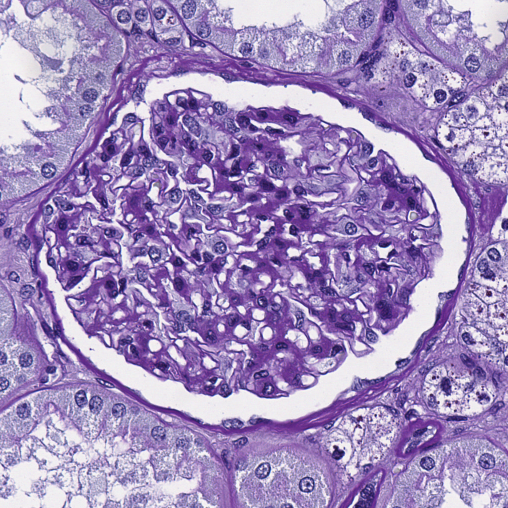}}
    \begin{tabular}{c c c c c c}
	& \small{PAS} & \small{Jones H\&E} & \small{CD68} & \small{Sirius Red} & \small{CD34}\\
    \rowname{\footnotesize{Original}} &
    \includegraphics[width=1.5cm]{images/cyclegan_images/02.png} &
    \includegraphics[width=1.5cm]{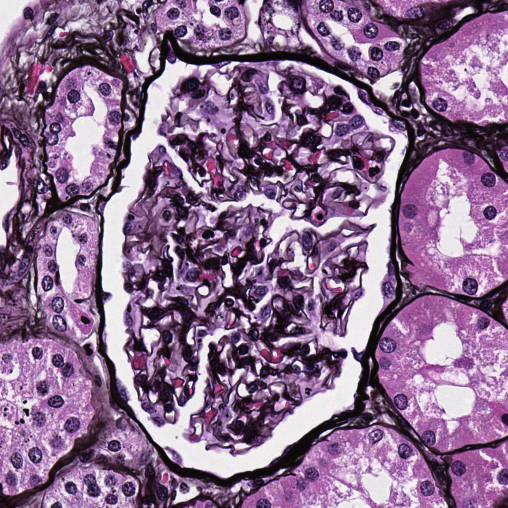} &
    \includegraphics[width=1.5cm]{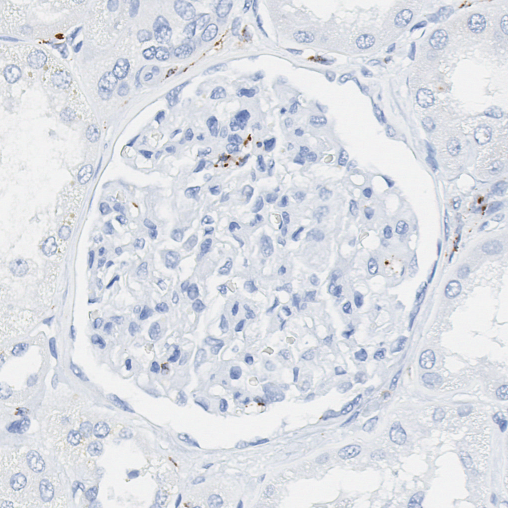} &
    \includegraphics[width=1.5cm]{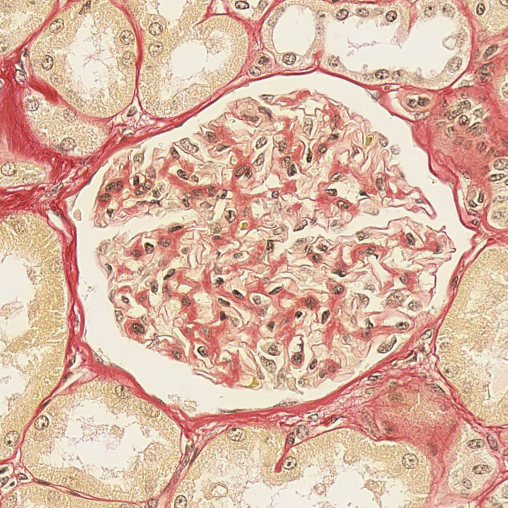} &
    \includegraphics[width=1.5cm]{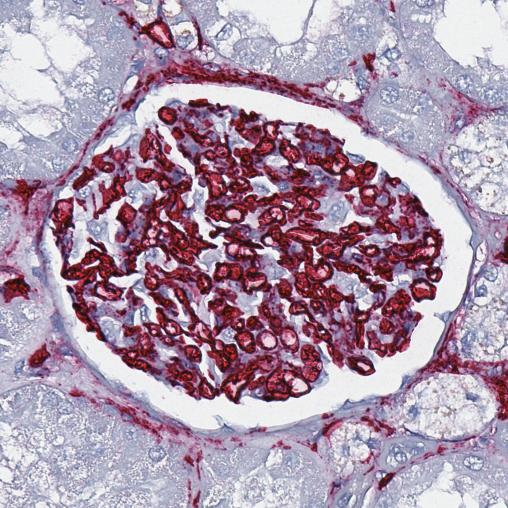}\\
    
    \multicolumn{2}{c}{\footnotesize{Target$\rightarrow$PAS}} & 
    \includegraphics[align=c,width=1.5cm]{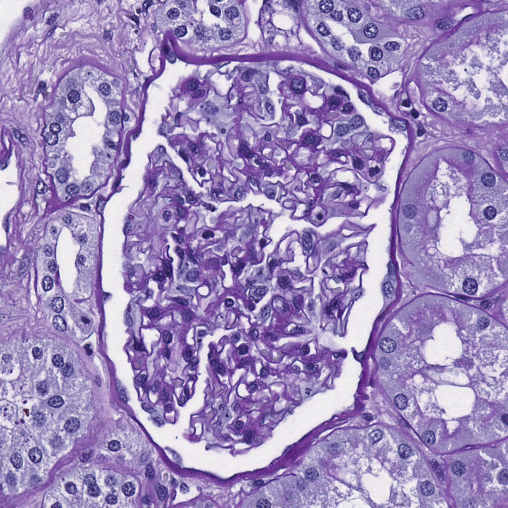} &
    \includegraphics[align=c,width=1.5cm]{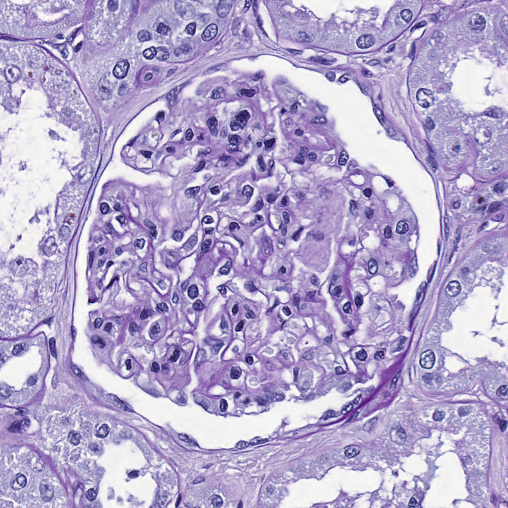} &
    \includegraphics[align=c,width=1.5cm]{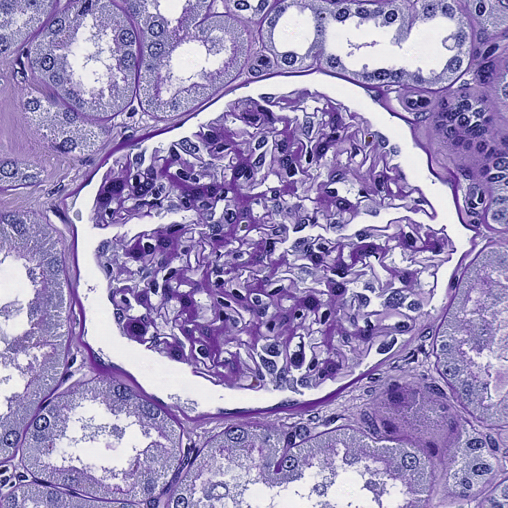} &
    \includegraphics[align=c,width=1.5cm]{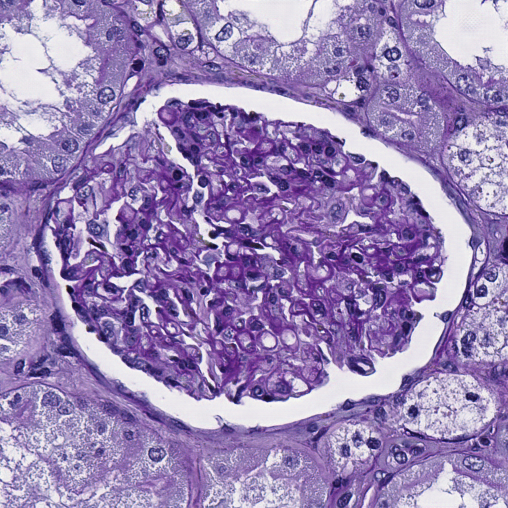}\\
    
    \addlinespace[0.1cm]
    \multicolumn{2}{c}{\footnotesize{PAS $\rightarrow$ Target}} & 
    \includegraphics[align=c,width=1.5cm]{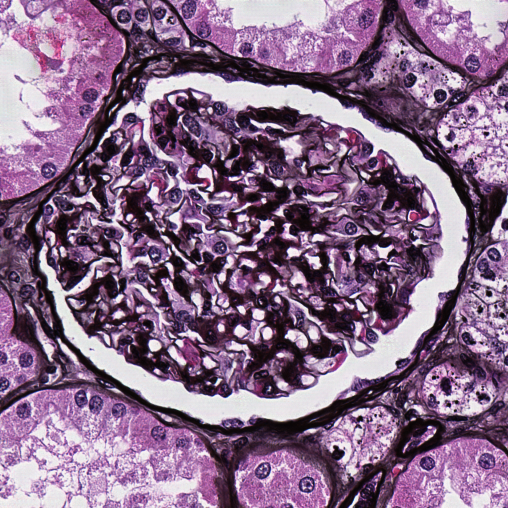} &
    \includegraphics[align=c,width=1.5cm]{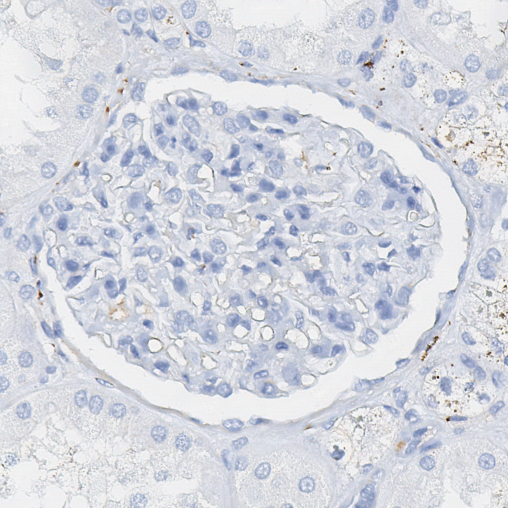} &
    \includegraphics[align=c,width=1.5cm]{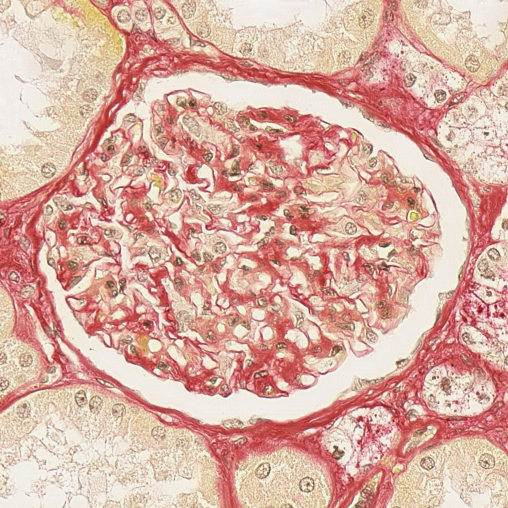} &
    \includegraphics[align=c,width=1.5cm]{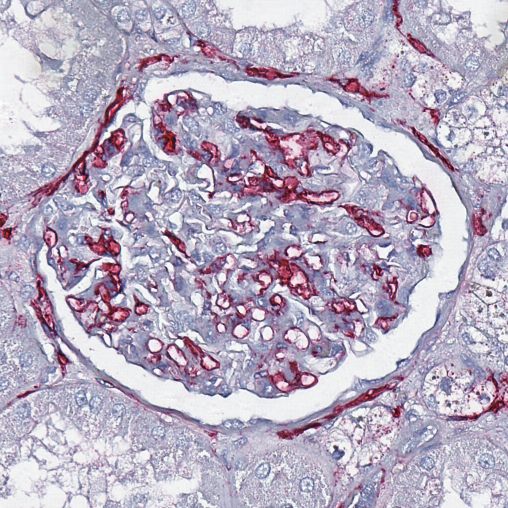}\\
    
    \end{tabular}
    \caption {Stain translations using CycleGAN.}
	\label{fig:cyclegan_translations}
	}
\end{figure}

Although visually these unpaired translations look very realistic, see Fig.\ \ref{fig:cyclegan_translations} (\nth{2} and \nth{3} row), in accordance with recent advances in the theoretical understanding of CycleGANs \cite{zee17bashkirova2019adversarial}, when translating from an information rich domain to an information poor domain some hidden information is embedded within them as imperceptible noise \cite{zee24vasiljevic2021self,zee1vasiljevic2021towards}. This can cause domain shift in the translated stains that can affect the final predictions, see Table \ref{table:MDS1} (\nth{2} row: CD68 $\rightarrow$ PAS and CD34 $\rightarrow$ PAS), since a majority of state-of-the-art computer vision algorithms are vulnerable to domain shift \cite{zee18csurka2017comprehensive}.

\section{Methods}\label{sec:methods}

\subsection{PixelCNN}
 \textcite{zee26song2017pixeldefend} have shown that PixelCNNs can be used to detect adversarial attacks in images, and we hypothesise that the hidden information may be detectable in a similar manner.
 
The PixelCNN \cite{zee19salimans2017pixelcnn++} is a generative model built specifically for images and to have tractable likelihood calculation. The model quantifies the pixels of an image $x$ over all its sub-pixels as a product of conditional distributions, such that
\begin{equation}\label{eq:pixelcnn}
	p(x) = \prod_{i=1}^{n^2} p(x_i|x_1, \dots, x_{i-1}).
\end{equation}
These conditional distributions are parameterised by a CNN and hence shared across all pixels in the image. PixelCNN is used to model the underlying data distribution of the source (i.e.\ PAS) and the translated Target$\rightarrow$PAS stains. %It is later leveraged to form a statistical test to conclude whether two distributions are equal.

\subsection{Domain Shift Metric}
The Domain Shift Metric (DSM) \cite{zee3de2019stain} measures the difference between two domains' distributions using the feature representations of a pre-trained model (referred to herein as Domain Shift Scores or DSS). Consider a CNN with layers $\{l_1,\dots,l_L\}$. Let $\Phi(x) = \{\phi_{l1}(x), \dots, \phi_{lk}(x)\}$ such that $\Phi_{lk}(x) \in \{\mathbb{R}^{h \times w}\}$ denote the filter activations at layer $l$ and filter $k$. The mean value of each $\Phi_{lk}(x)$ is calculated as
\begin{equation} c_{lk}(x) = %\left( 
\frac{1}{h w} 
%\right) 
\sum_{i,j}^{h,w} \Phi_{lk}(x)_{i,j}.
\label{eq:domain_shift_part1}
\end{equation}
Let $p^\mathcal{S}_{c_{lk}}(x)$ denotes continuous distribution of $c_{lk}(x)$ over the source stain $\mathcal{S}$ and $p^\mathcal{T}_{c_{lk}}(x)$ denotes the same over the translated Target$\rightarrow$PAS stain $\mathcal{T}$, then the DSM $R_l$ is defined as
\begin{equation} 
R_l(p^\mathcal{S}, p^\mathcal{T}) = \frac{1}{k} \sum_{i=1}^{k} \mathcal{D} \left( p^\mathcal{S}_{c_{lk}}, p^\mathcal{T}_{c_{lk}} \right),
\label{eq:domain_shift_part2}
\end{equation}
where $\mathcal{D}$ is the Wasserstein distance \cite{zee27ramdas2017wasserstein} between $p^\mathcal{S}_{c_{lk}}(x)$ and $p^\mathcal{T}_{c_{lk}}(x)$, which tends towards zero when $\mathcal{S}$ and $\mathcal{T}$ are similar. 
%\color{red} Tom here I think we should made our code publicly available for domain shift measure. What are your thoughts about it?

\color{black} 
\section{Experiments and Results}
\label{sec:experiments_and_results}

\begin{figure*}[]
\setlength{\tabcolsep}{0pt}
\captionsetup{}
    \begin{tabular}{c c c c}
    \includegraphics[width=0.25\textwidth]{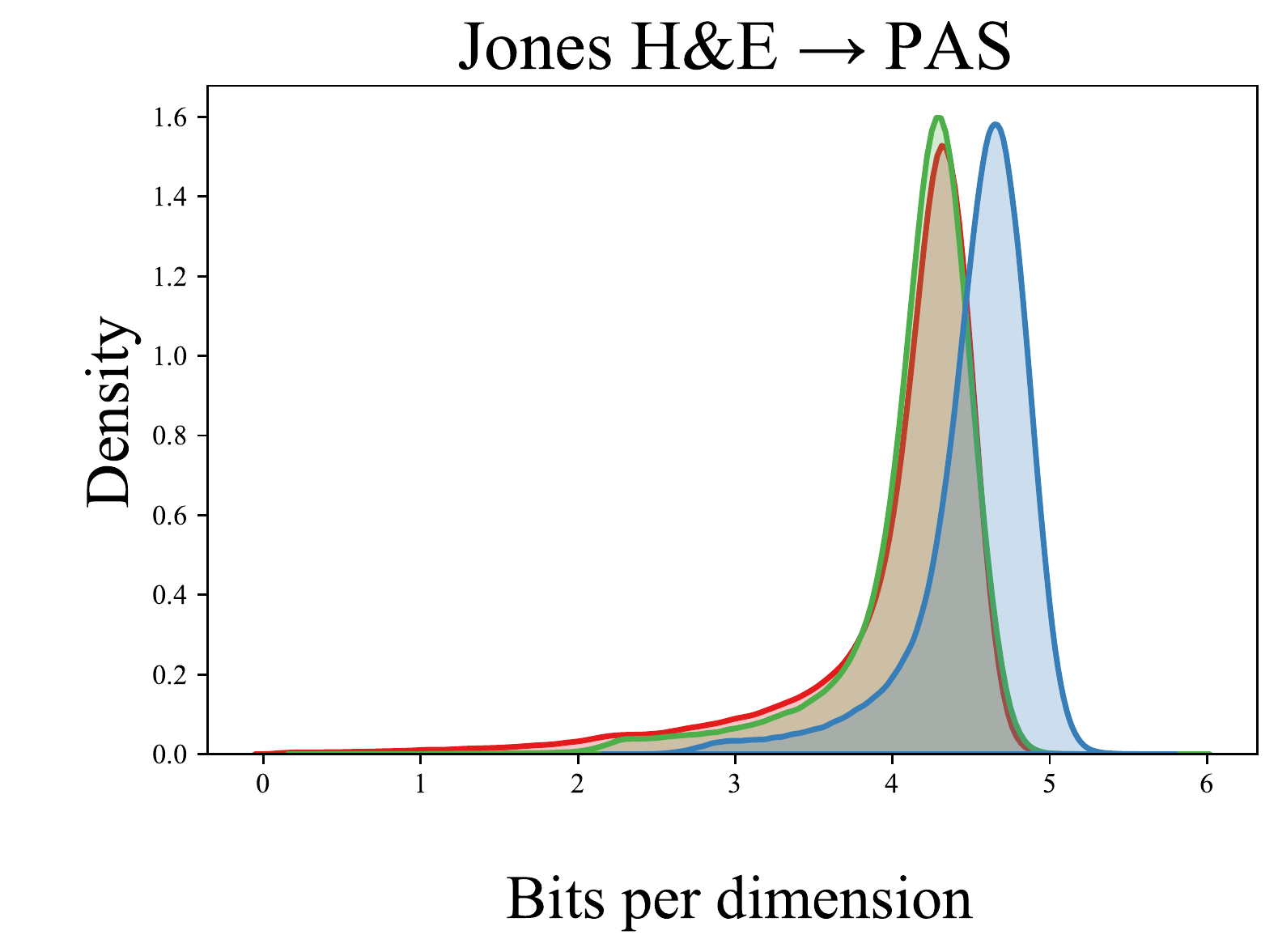} & 
    \includegraphics[width=0.25\textwidth]{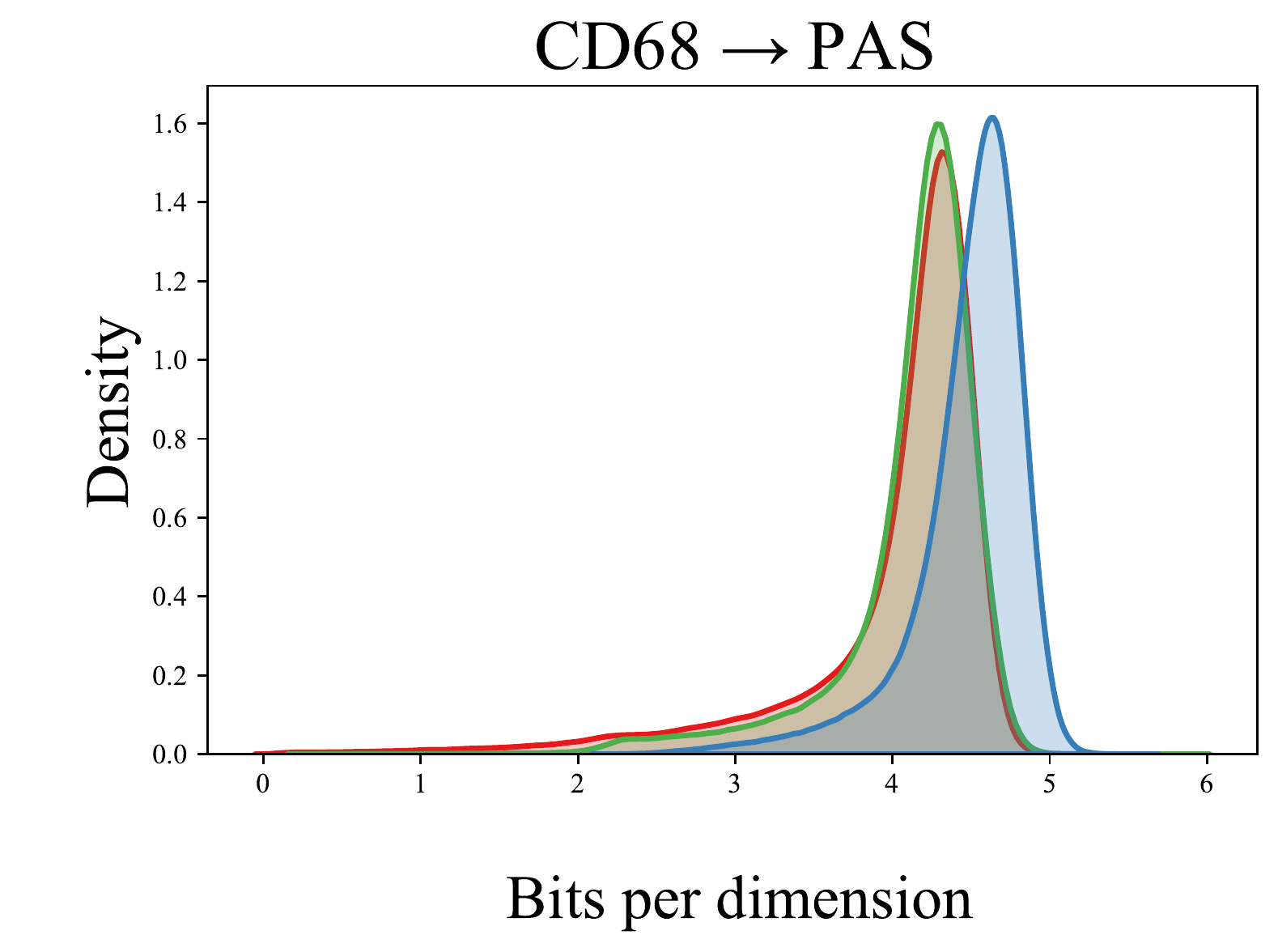} &
    \includegraphics[width=0.25\textwidth]{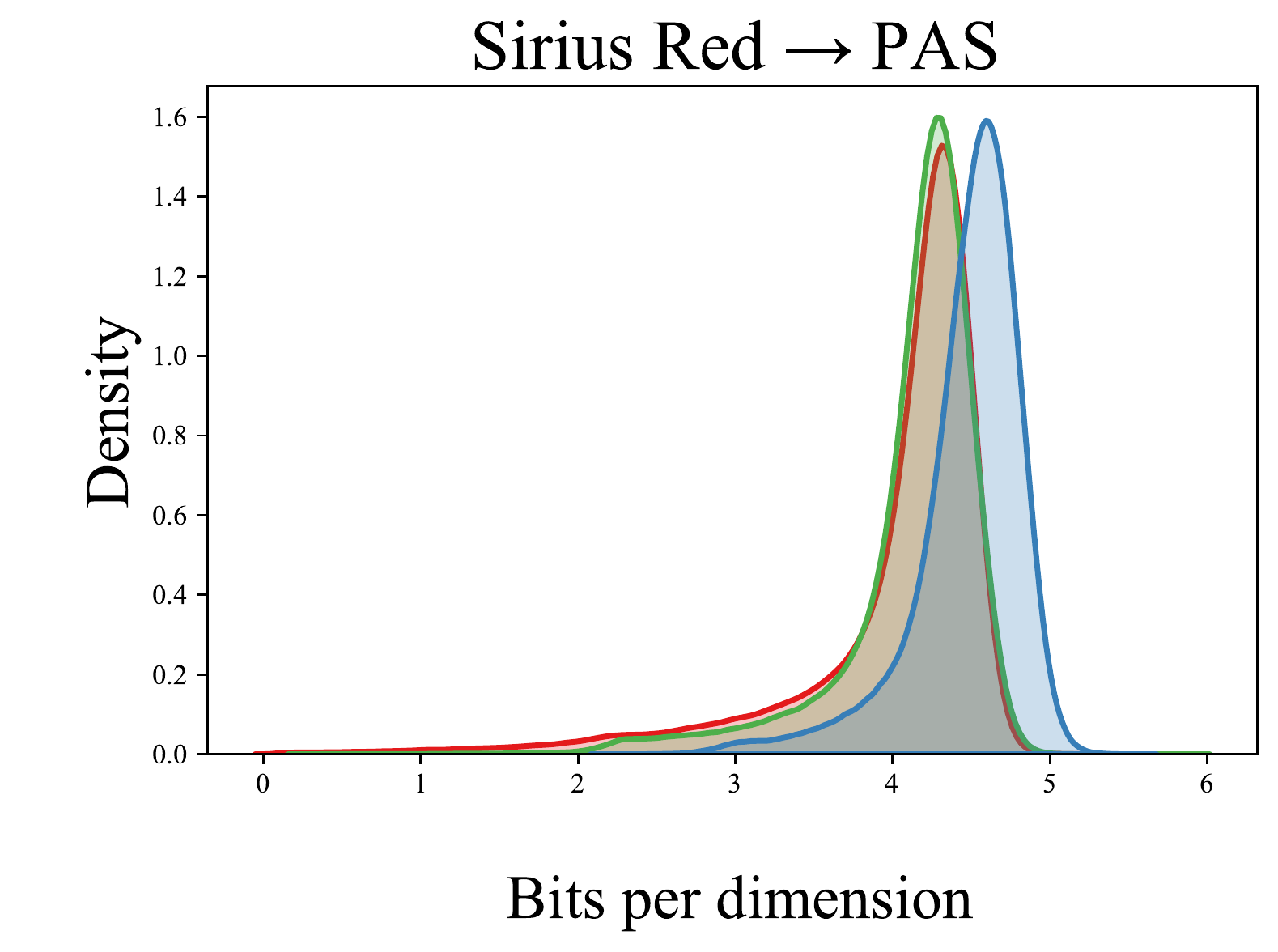} & 
    \includegraphics[width=0.25\textwidth]{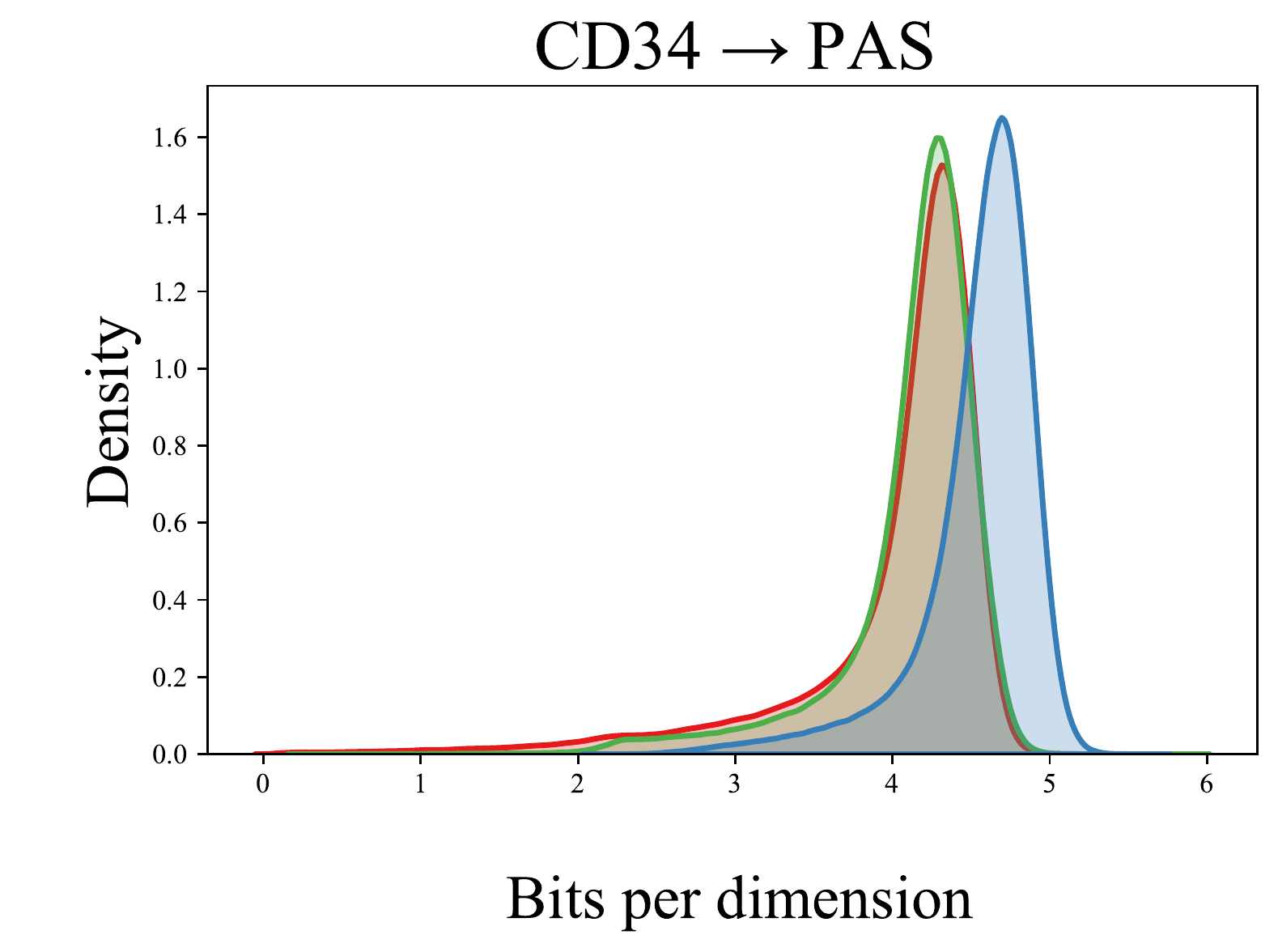} \\
    
    \hspace{10mm}\scriptsize{Wasserstain Distance = 0.5371} & \hspace{10mm}\scriptsize{Wasserstain Distance = 0.4935} & \hspace{10mm}\scriptsize{Wasserstain Distance = 0.4816} & \hspace{10mm}\scriptsize{Wasserstain Distance = 0.5801}\\
    \multicolumn{4}{c}{\includegraphics[width=0.5\textwidth]{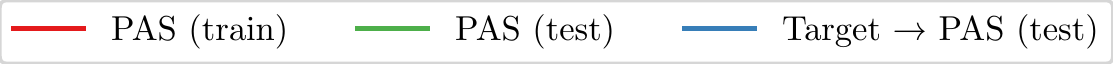}}\\
    \end{tabular}
    \caption{PixelCNN based domain shift measures of the translated Target$\rightarrow$PAS stains w.r.t.\ real PAS train and test sets.} 
    \label{fig:pixelcnn_distributions}
\end{figure*}

\subsection{Data}
Tissue samples were obtained from a group of 10 patients who had tumor nephrectomy for renal carcinoma. Renal tissue was chosen to be as far from the tumors as possible to represent
largely normal renal glomeruli; however, certain samples contained varying degrees of pathological modifications such as complete or partial displacement of functional tissue by fibrotic changes (``scerosis'') indicating normal age-related changes or the renal effects of general cardiovascular comorbidity (e.g.\ cardial arrhythmia, hypertension, arteriosclerosis). Using an automated staining tool (Ventana Benchmark Ultra), the paraffin-embedded samples were sliced into 3µm thick sections and stained with either Jones H\&E basement membrane stain (Jones), PAS, Sirius Red, as well as two immunohistochemistry markers (CD34, and CD68). An Aperio AT2 scanner was used to capture whole slide pictures at 40 magnification (a resolution of 0.253 m / pixel). Pathology specialists annotated and verified all of the glomeruli in each WSI by labeling them with Cytomine \cite{zee28maree2016collaborative}. The whole dataset was split into 4 training, 2 validation, and 4 test patients. The number of glomeruli in each staining dataset was: PAS - 662 (train), 588 (valid), 1092 (test); Jones H\&E - 590 (valid), 1043 (test); Sirius Red - 576 (valid), 1049 (test); CD34 - 595 (valid), 1019 (test); CD68 - 521 (valid), 1046 (test). 
\subsection{Experiments}
Throughout the experiments, patches of size 508$\times$508 pixels are used since glomeruli and part of the surrounding area fit within this patch size at the level-of-detail used. %PAS is assumed to be the source stain for which we have annotations available and the other stains as target for which annotations do not exist.% 
To account for random variations, all experiments are repeated with three different CycleGAN models and each is used with five different U-Net models, i.e.\ 15 total repetitions. When training (PixelCNN \& CycleGAN) 5000 training and 500 validation patches were used, and when evaluating (PixelCNN \& DSM) 5 sets of 1000 test patches  were used. These were extracted in a random, uniform manner from the corresponding patients.

\subsubsection{U-Net}
The U-Net \cite{zee21ronneberger2015u} architecture is used to segment the glomerulus region in the source staining (PAS), since it has been proven successful in biomedical imaging \cite{zee22litjens2017survey} and, in particular, glomeruli detection \cite{zee23de2018automatic}. Glomeruli segmentation is framed as a two class problem: glomeruli (pixels that belong to glomerulus), and tissue (pixels outside a glomerulus). The training set comprised all glomeruli from the source stain (PAS) training patients (662) plus 4634 tissue (i.e.\ non-glomeruli) patches (accounting for the variance in non-glomeruli tissue). The network was trained using the same parameters and procedures as used by \textcite{zee10lampert2019strategies}. The segmentation scores for each stain are presented in the \nth{1} row of Table \ref{table:MDS1}. % The network was trained with these parameters: batch\_size : 8, learning\_rate : 0.0001, epochs : 250. During pre-processing we removed the slide background (non-tissue) by having a threshold at each image by its mean value and then by removing small object closing holes. All patches are standardised to [0, 1] and normalised by the mean and standard deviation of the (labeled) training set.  The following augmentations are applied with an independent probability of 0.5 (batches are augmented "on the fly"), in order to further force the network to learn general features: elastic deformation ($\sigma$ = 10, $\alpha$ = 100); random rotation in the range [0°, 180°], and horizontal/vertical flip; additive Gaussian noise with $\sigma \epsilon$ [0, 0.01]; Gaussian filtering with $\sigma \epsilon$ [0, 1]; brightness, colour, and contrast enhancements with factors sampled from [0.9, 1.1]; stain variation by colour. The model with the lowest validation loss is kept.

\subsubsection{CycleGAN}
CycleGAN \cite{zee16zhu2017unpaired} is an unpaired image-to-image translation network widely used for style transfer in digital pathology \cite{zee1vasiljevic2021towards, zee11gadermayr2019generative, zee24vasiljevic2021self}. Given images of a source stain $s \sim \mathcal{S}$ and a target stain $t \sim \mathcal{T}$, the goal is to learn a two way mapping between $t$ and $s$. %, such that
%\begin{equation*}
%\label{equation:cyclegan}
%\begin{multlined}
%\mathcal{L}_{cyc}(G_1, G_2) = \mathbb{E}_s [\|s - G_2(G_1(s))\|_1] + \\
%\mathbb{E}_t [\|y - G_1(G_2(t))\|_1].
%\end{multlined}
%\end{equation*}
This network is used to translate all of the target stains (i.e.\ Jones H\&E, CD68, Sirius Red, and CD34) to the source stain (PAS). % as in Fig.\ \ref{fig:cyclegan_translations} (\nth{2} row).
\textcite{zee11gadermayr2019generative} briefly explain how different sampling strategies for the annotated and unannotated stains can negatively impact a stain transfer model’s performance, and therefore patches are randomly extracted in an unsupervised manner using a uniform sampling strategy. The same training strategies as used by \textcite{zee1vasiljevic2021towards} was employed for training the CycleGAN networks to translate the target stains to PAS. Fig.\ \ref{fig:cyclegan_translations} (\nth{2}, and \nth{3} row) present the results for each of these translations and the second row in Table \ref{table:MDS1_domain_shift} presents the U-Net segmentation scores when the target stains are translated to PAS and the PAS trained network used to segment them.
% \color{red} \textcite{zee11gadermayr2019generative} showed that different data sampling techniques for the annotated and unannotated domains might have a negative influence on the performance of a stain transfer model, thus patches are extracted in a random manner using a uniform sampling strategy (i.e.\ in an unsupervised manner). 
\color{black}

\begin{figure}[]
    \centering
    % \begin{subfigure}[b]{0.425\textwidth}
    % \includegraphics[width=\textwidth]{images/scatter_plots/mean_correlation_scatterPlot_2000_samples_randomly_shuffled.pdf}
    % \caption{}
    % \label{fig:scatter_plots_a}
    % \end{subfigure}
    \includegraphics[width=0.5\textwidth, center]{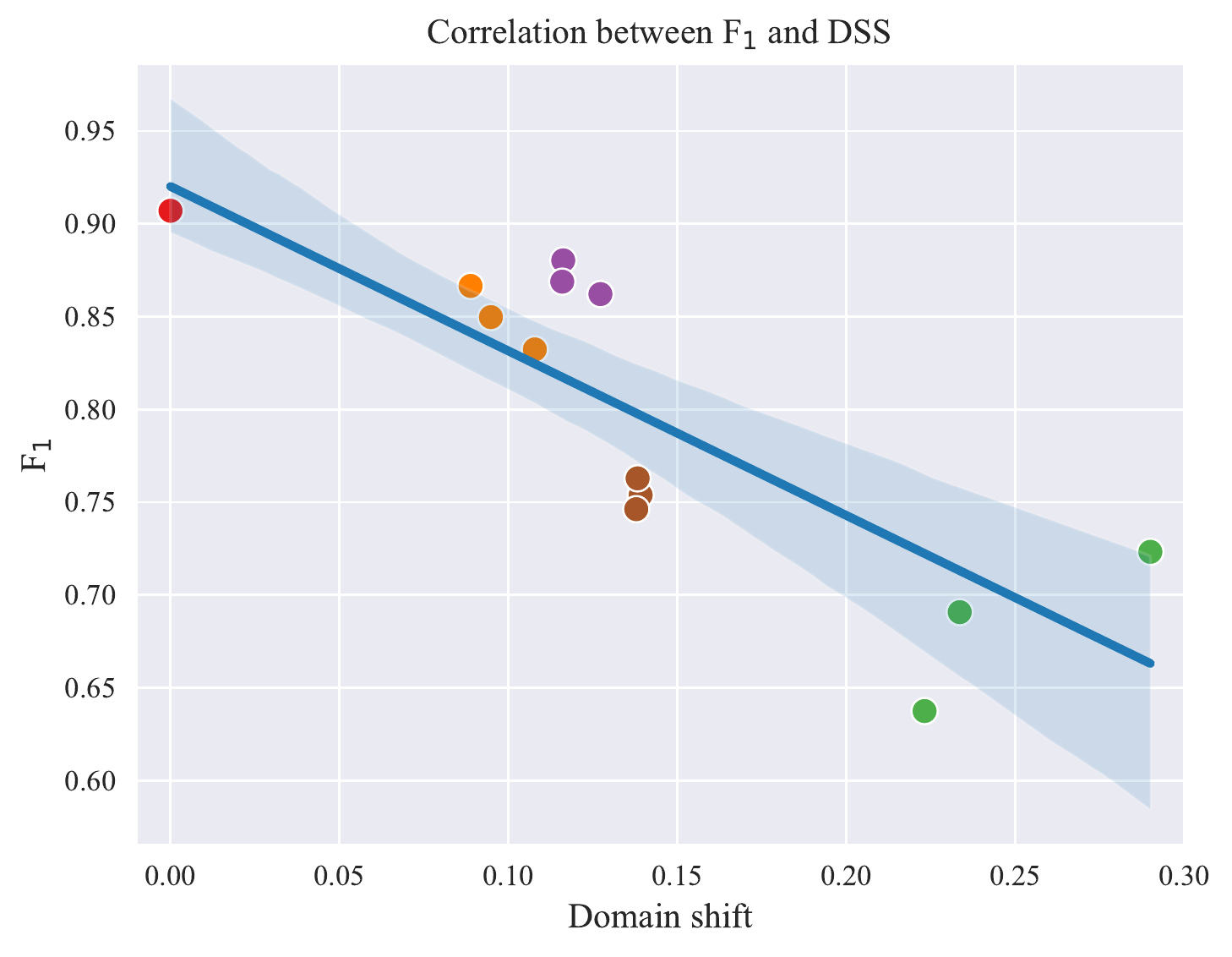}\\
    \includegraphics[width=0.48\textwidth, right]{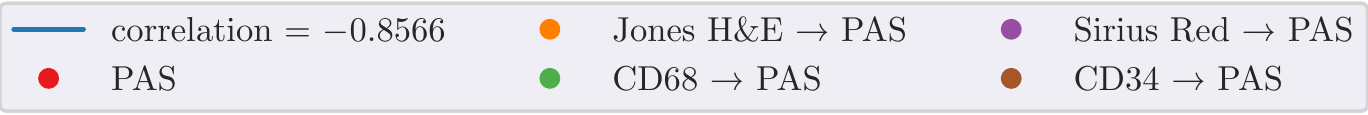}
    \caption{Correlation between segmentation scores 
    % and DSSs for (a) 1000 glomeruli and 1000 negative test patches translated to PAS (b) segmentation
    of the whole test slides translated to PAS and the average DSS of 5 sets of 1000 randomly sampled test patches. Each point is the average of 5 U-Net repetitions for each CycleGAN model.} 
    \label{fig:scatter_plots}
\end{figure}

\subsubsection{PixelCNN}
%PixelCNN \cite{zee19salimans2017pixelcnn++} is a generative model that can model the underlying distribution of image data. 
%Although the stain translations (Target$\rightarrow$PAS) look very realistic and glomeruli retains its morphological structure in the translations but still the segmentation model trained on PAS is experiencing a significance drop in its performance.

As was shown in Table \ref{table:MDS1} (\nth{2} row), a segmentation model trained on PAS experiences a degradation in performance when applied to translated images. It is our hypothesis that this is caused by the imperceptible noise added during stain style transfer. This noise is imperceptible to humans but causes a domain/covariate shift. To test this, the probability density of the underlying source (PAS) training distribution is estimated using a PixelCNN model. 
%We employ PixelCNN here because of its state of art performance in modeling image distribution. 

As each pixel value is conditioned on the product of all previously generated pixels, the original architectures were trained and evaluated on patches of size $32\times32$ due to GPU memory limitations. 
%These small resolution patches were extracted from the original $508 \times 508$ on the fly. 
Therefore 5000 (training \& test) and 500 (validation) patches were decomposed into \num[group-separator={,}]{1280000} non-overlapping training and test and \num[group-separator={,}]{128000} validation patches. 
%were randomly extracted using a uniform sampling strategy and 
The same training parameters as used by \textcite{zee19salimans2017pixelcnn++}\footnote{\url{https://github.com/openai/pixel-cnn}} were employed. The training of one PixelCNN model took approximately 15 days using four V100 GPUs (in parallel).

%A trained PixelCNN generative model is very sensitive to imperceptible noise in the case of adversarial attacks \cite{zee26song2017pixeldefend}. 
The PixelCNN model is first validated using the PAS training data and an unseen PAS test set. It is found that their log-likelihoods follow the same order of magnitudes, see Fig.\ \ref{fig:pixelcnn_distributions}.
The log-likelihood distributions of the Target$\rightarrow$PAS stains are also included in this figure and they clearly show that there is a domain shift compared to the overlapping test distributions.
%compared to real PAS train set is presented in Fig.\ \ref{fig:pixelcnn_distributions}. 
%Besides qualitative analysis, later with these likelihood magnitudes we provide a quantitative measure to detect the existence of domain shift. 
The Wasserstein distance can be used to measure the similarity of two distributions, where smaller distances indicate more similar distributions. This is the case for the train (PAS) and test (PAS) distributions, giving a Wasserstein distance of $0.0879$ (average over 5 sets of 1000 randomly sampled patches). In comparison, the distance between PAS train and Target$\rightarrow$PAS for all stains is relatively large, see Fig.\ \ref{fig:pixelcnn_distributions}, highlighting a greater difference between the distributions.

There is a strong correlation, $-0.7339$, between full slide segmentation scores (Table \ref{table:MDS1}: \nth{2} row) and distribution distance. This unsupervised approach can therefore be used for insight into the success of applying the segmentation model to unseen, unannotated data that has undergone domain shift by using a sample of the data. Furthermore, it is also able to detect 
%The PixelCNN is an unsupervised method that is powerful enough to detect 
imperceptible (even to domain experts) domain shifts.
%nevertheless, a segmentation network may be affected by such shifts in a non-linear way. It can, however, be used to determine whether a set of translations match the original distribution or not.

%that is even barely perceptible to humans or even  to domain experts. Therefore it has no proper direction to correlate with segmentation accuracy. To show a perfect correlation of domain shift and segmentation scores, we employed a newly introduced domain shift metric in section \ref{sec:domain_shift_metric}. 

% \color{black}
% % The Kolmogorov-Smirnov (KS
% ) test \cite{zee25hodges1958significance} tests the null hypothesis $H_o$ that the two distributions are identical and therefore an alternative hypothesis $H_a$ that they are not and can be used to measure this quantitatively. If the KS statistic between two distributions is small then the null hypothesis cannot be rejected \cite{zee25hodges1958significance}, which is the case for the train (PAS) and test (PAS) distributions at a confidence of \color{red} ???? \color{black}, see Table \ref{table:pixelcnn_kstest}, i.e.\ they are identical.  
%is so small and so we can not reject our null hypothesis 
% For all the translated distributions Target$\rightarrow$PAS the KS statistics 
%between train (PAS) and test stains of target translated to PAS is not small and hence we can 
% do not lead to the rejection of the %reject our 
% null hypothesis,
%and can go in the favour of alternative hypothesis (
% i.e.\ these distributions are not identical.
%), see Table \ref{table:pixelcnn_kstest}.

\color{black}
\subsubsection{Domain Shift Metric} \label{sec:domain_shift_metric}

\begin{table}[]
\centering
%\footnotesize
\captionsetup{}%P{2.2cm} 
     \begin{tabular}{ 
     P{1.7cm} P{1.7cm} P{1.7cm} P{1.7cm}}
      \hline 
      %\multirow{3}{*}{\shortstack{Sampling\\ Strategy}} & 
      %\multicolumn{4}{c}{Test Stain}\\
      \multirow{2}{*}{\shortstack{Jones \\H\&E $\rightarrow$PAS}} & \multirow{2}{*}{CD68$\rightarrow$PAS} & \multirow{2}{*}{\shortstack{Sirius \\Red$\rightarrow$PAS}} & \multirow{2}{*}{CD34$\rightarrow$PAS}\\
      & & & \\
      \hline
      \hline
    %   \multirow{2}{*}{%\shortstack{MDS1 \\ %\footnotesize{Target$\rightarrow$PAS} \\
    %   {2000 patches}}%} %&
    %   %\multirow{2}{*}{\shortstack{F$_1$}} & \multirow{2}{*}{\shortstack{0.934 \\ \small{(0.0008)}}} & \multirow{2}{*}{\shortstack{0.713 \\ \small{(0.061)}}} & \multirow{2}{*}{\shortstack{0.944 \\ \small{(0.003)}}} & \multirow{2}{*}{\shortstack{0.788 \\ \small{(0.009)}}}\\
    %   %& & & & & \\
    %   & \multirow{2}{*}{\shortstack{0.089 \\ \scriptsize{(0.004)}}} & \multirow{2}{*}{\shortstack{0.307 \\ \scriptsize{(0.010)}}} & \multirow{2}{*}{\shortstack{0.096 \\ \scriptsize{(0.009)}}} & \multirow{2}{*}{\shortstack{0.240 \\ \scriptsize{(0.008)}}}\\
    %   & & & & \\
    %   \hline
      %\multirow{2}{*}{%\shortstack{MDS1 \\ %\footnotesize{Target$\rightarrow$PAS} \\ 
      %{Random Patches}}%} %&  %\multirow{2}{*}{\shortstack{F$_1$}} & \multirow{2}{*}{\shortstack{0.849 \\ \small{(0.017)}}} & \multirow{2}{*}{\shortstack{0.683 \\ \small{(0.043)}}} & \multirow{2}{*}{\shortstack{0.870 \\ \small{(0.009)}}} & \multirow{2}{*}{\shortstack{0.7543 \\ \small{(0.008)}}}\\
      %& & & & & \\
       %& 
       \multirow{2}{*}{\shortstack{0.097 \\ \scriptsize{(0.008)}}} & \multirow{2}{*}{\shortstack{0.248 \\ \scriptsize{(0.002)}}} & \multirow{2}{*}{\shortstack{0.119 \\ \scriptsize{(0.003)}}} & \multirow{2}{*}{\shortstack{0.138 \\ \scriptsize{(0.002)}}}\\
      & & & \\
      \hline
     \end{tabular}
     \caption{Average Domain Shift Scores (R$_l$) of 5 sets of 1000 randomly sampled patches for the Target$\rightarrow$PAS translated stains. Averages of 5 U-Net and 3 CycleGAN repetitions, i.e.\ 15 repetitions in total; standard deviations are in parentheses.
     %Larger the F$_1$ score smaller will be the $R_l$ distance between two distributions.
     }
    \label{table:MDS1_domain_shift}
\end{table}

Using the pre-trained PAS source network as the feature representations, the domain shift can also be calculated using the domain shift metric, Eq.\ \eqref{eq:domain_shift_part2}. The DSS are presented in Table \ref{table:MDS1_domain_shift}. Since the model is supervised on the same task, the average segmentation score (of 5 models) has a stronger negative correlation than observed with the PixelCNN, as in Fig.\ \ref{fig:scatter_plots}.

%Here the 5000 samples used to calculate DSS are split into five sets of 1000 samples and the average DSS presented, see Table
 
%Giving an even stronger method for inferring the performance of the model on an unseen, unannotated sample.

%To test the generality and real-world validity of this approach, the DSS is calculated in the same manner as PixelCNN (except that the 5000 samples used to calculate DSS are split into five sets of 1000 samples and the average DSS presented), see Table \ref{table:MDS1_domain_shift}. 
%The correlation of these with the F$_1$ score of the full test slides is presented in 
%The results this far have been calculated with a balanced dataset created to have segmentation and domain shifts for exactly the same data. The same experiment is repeated with the segmentation performance of the full test slides, using the domain shift calculated on ???? random patches extracted from the test slides, which represents the real-world application of such an approach. 
%Fig.\ \ref{fig:scatter_plots_b}, confirming that the same trend is observed (although with a slightly weaker correlation) and that 

This stronger correlation is observed when compared to the PixelCNN since DSM uses the same representation as the segmentation model, which has been trained in a supervised manner for a specific task. It is therefore sensitive to the type of domain shift that will affect the segmentation performance.\color{black}
%where large values of segmentation scores (F$_1$) are observing a less shift in domain.  
%It is an obvious claim here that if smaller is the domain shift distance observed between $\mathcal{S}$ and $\mathcal{T}$ stains larger will be the segmentation scores and if larger is the domain shift distance observed smaller will be the segmentation scores. 

%\color{red} 
%\begin{figure}[!ht]
%\captionsetup{}
%    \includegraphics[width=0.5\textwidth]{images/scatter_plots/mean_scores_different_samples.pdf}
%    \caption{} 
%    \label{fig:line_plot}
%\end{figure}
%I have to write a paragraph about the relationship between number of samples and domain shift scores.    

%\color{black}
\section{Conclusions} \label{sec:conclusion_future}

This article has investigated unsupervised approaches to propose a method to detect domain shift in histopathological images and shown that domain shift has a strong correlation with the segmentation performance of stain translated data. 
%The proposed solution is general and is applicable to all computer vision applications. 
As such, the work focused on detecting imperceptible noise that is introduced by CycleGAN models, however the solution is general and can detect any kind of domain shift.  %The domain shift can be estimated both visually and quantitatively, and it also 

These measures offer a mechanism to estimate the average performance of pre-trained neural networks when applied to unseen target stains (for the same task) without having any expert opinion or ground-truth. This has been achieved using two approaches, one that uses an unsupervised, generative model of the data and another that uses a pre-trained (supervised representation). Since the purpose of this work is to predict how domain shift will affect a pre-trained model, this representation would be available, however, if this is not the case, then the completely unsupervised PixelCNN also offers strong correlation with segmentation score.

%This work will be extended to form a loss function for translation models to limit domain shift and to learn domain invariant representations.

% References should be produced using the bibtex program from suitable
% BiBTeX files (here: strings, refs, manuals). The IEEEbib.bst bibliography
% style file from IEEE produces unsorted bibliography list.
% ------------------------------------------------------------------------- 

\section{Acknowledgments}
\label{sec:acknowledgments}

%!!!!!!!!!!!!!!!!!!!!!!!!!!!!!!!
% DO NOT USE THIS TEXT WHEN IT IS NOT ANONYMISED, USE THE COMMENTED TEXT BELOW...
%!!!!!!!!!!!!!!!!!!!!!!!!!!!!!!!
This work was supported by: ArtIC project ''Artificial Intelligence for Care'' (grant ANR-20-THIA-0006-01) and co-funded by Région Grand Est, Inria Nancy - Grand Est, IHU of Strasbourg, University of Strasbourg and University of Haute-Alsace; ERACoSysMed and e:Med initiatives by BMBF; SysMIFTA (project management PTJ, FKZ 031L-0085A; ANR, project number ANR-15-CMED-0004); and the French Government for co-tutelle funding (Jelica Vasiljevic). We thank the Nvidia Corporation, the Centre de Calcul de l'Universite de Strasbourg, and GENCI-IDRIS (Grant 2020-A0091011872) for access to the GPUs used for this research. We also thank the Medizinische Hochschule Hanover for providing high-quality images and annotations.

%!!!!!!!!!!!!!!!!!!!!!!!!!!!!!!!
% USE THIS VERSION WHEN NOT ANONYMISED
%!!!!!!!!!!!!!!!!!!!!!!!!!!!!!!!
%This work was supported by: ANR/University of Strasbourg (Zeeshan Nisar); ERACoSysMed and e:Med initiatives by BMBF; SysMIFTA (project management PTJ, FKZ 031L-0085A; ANR, project number ANR-15-CMED-0004); SYSIMIT (project management DLR, FKZ 01ZX1608A); and the French Government for co-tutelle funding (Jelica Vasiljevi\'{c}). We thank the Nvidia Corporation, the \emph{Centre de Calcul de l'Universit\'{e} de Strasbourg}, and GENCI-IDRIS (Grant 2020-A0091011872) for access to the GPUs used for this research. We also thank the \emph{Medizinische Hochschule Hanover} for providing high-quality images and annotations.

\section{Compliance with Ethical Standards}
%\color{red} WE CAN REMOVE THIS IF NECESSARY\color{black}

Study performed in line with the principles of the Declaration of Helsinki. Approval granted by the Ethics Committee of Hanover Medical School (Date 12/07/2015, No.\ 2968-2015).

\printbibliography

%\bibliographystyle{IEEEbib}
%\bibliography{strings}

\end{document}